\documentclass[11pt,draftclsnofoot,peerreviewca,letterpaper,onecolumn]{IEEEtran}

\usepackage{cite}
\usepackage{graphicx,color,epsfig,rotating}
\usepackage{amsfonts,amsmath,amssymb,bbm,bm}
\usepackage{algorithm}
\usepackage{algpseudocode}
\usepackage{subfigure}
\usepackage{amsmath}
\usepackage{cite}
\usepackage{mdwtab}
\usepackage{placeins}
\usepackage{psfrag, graphicx}
\usepackage[latin1]{inputenc}
\usepackage{amssymb}
\usepackage{multirow}
\usepackage[shortlabels]{enumitem}

\ifodd 1

\else

\fi

\newtheorem{defn}{Definition}
\newtheorem{theorem}{Theorem}
\newtheorem{lemma}{Lemma}
\newtheorem{corollary}{Corollary}
\newtheorem{claim}{Claim}
\newtheorem{remark}{Remark}

\newcommand*{\rom}[1]{\expandafter\@slowromancap\romannumeral #1@}
\usepackage{changes}
\definechangesauthor[color=red,name={Mingyue Ji}]{MJ}

\long\def\comment#1{}

\newfont{\bbb}{msbm10 scaled 700}

\newfont{\bb}{msbm10 scaled 1100}

\newcommand{\Nc}{{\cal N}}

\newcommand{\be}{\begin{equation}}
\newcommand{\ee}{\end{equation}}
\newcommand{\bea}{\begin{eqnarray}}
\newcommand{\eea}{\end{eqnarray}}

\begin{document}

\sloppy


\title{Private Information Retrieval from Heterogeneous Uncoded Storage Constrained Databases with Reduced Sub-Messages}


\author{
    \IEEEauthorblockN{ Nicholas Woolsey,
		Rong-Rong Chen, and Mingyue Ji }

	\IEEEauthorblockA{Department of Electrical and Computer Engineering, University of Utah\\
		Salt Lake City, UT, USA\\
		Email: \{nicholas.woolsey@utah.edu,
		 rchen@ece.utah.edu,
		mingyue.ji@utah.edu\}}
}

\maketitle


\begin{abstract}
We propose capacity-achieving schemes for private information retrieval (PIR) from uncoded databases (DBs) with both homogeneous and heterogeneous storage constraints. In the PIR setting, a user queries a set of DBs to privately download a message, where privacy implies that no one DB can infer which message the user desires. In general, a PIR scheme is comprised of storage placement and delivery designs. Previous works have derived the capacity, or infimum download cost, of PIR with uncoded storage placement and also sufficient conditions of a storage placement design to meet capacity. However, the currently proposed storage placement designs require splitting each message into an exponential number of sub-messages with respect to the number of DBs. In this work, when DBs have the same storage constraint, we propose two simple storage placement designs that satisfy the capacity conditions. Then, for more general heterogeneous storage constraints, we translate the storage placement design process into a ``filling problem". We design an iterative algorithm to solve the filling problem where, in each iteration, messages are partitioned into sub-messages and stored at subsets of DBs. All of our proposed storage placement designs require a number of sub-messages per message at most equal to the number of DBs.

\end{abstract}
\section{Introduction}
\label{sec: Intro}
The private information retrieval (PIR) problem originally introduced by Chor \textit{et al.} \cite{chor1995private, chor1998private} has been recently studied under an information theoretic point of view \cite{sun2017capacity}. In the PIR problem, a user privately downloads one of $K$ messages from a set of $N$ non-colluding databases (DBs). Moreover, privacy implies that no DB can infer which of the $K$ messages the user is downloading. To achieve privacy the user generates strategic queries to the DBs such that sub-messages from all $K$ messages are requested. To gauge the performance of the PIR scheme, the rate, $R$, is defined as the ratio of desired bits (or size of each message), $L$, to the total number of downloaded bits, $D$. In the traditional setting of full storage PIR (FS-PIR), each DB has access to all $K$ messages and the capacity, or maximum achievable rate, of PIR is $\left( 1+ \frac{1}{N} + \frac{1}{N^2} + \cdots + \frac{1}{N^{K-1}} \right)^{-1}$ \cite{sun2017capacity}. Multiple achievable schemes have been developed which achieve FS-PIR capacity by exploiting downloaded undesired sub-messages for coding opportunities \cite{sun2017capacity,tian2018capacity,sun2017optimal}.

More recently, the problem of homogeneous storage constrained PIR (SC-PIR) was proposed such that each DB can only store $\mu KL$ bits 
where $\frac{1}{N} \leq \mu \leq 1$ \cite{tandon2018pir}.  
Define $t = \mu N$, the capacity of homogeneous SC-PIR was shown to be the convex hull of the points $\left( 1+ \frac{1}{t} + \frac{1}{t^2} + \cdots + \frac{1}{t^{K-1}} \right)^{-1}$ for $t = 1,2.\ldots,N$ \cite{wei2018capacity,attia2018capacity}. Different from FS-PIR, there is an additional design aspect to SC-PIR which is the storage placement must be strategically designed. For example, the original homogeneous SC-PIR scheme met capacity \cite{tandon2018pir} by using the storage placement scheme of the classical shared link coded caching problem \cite{maddah2014fundamental}. One of the limitations of this scheme is the storage placement requires that each message is split into $O(\exp N)$ sub-messages. 
Hence, the proposed PIR scheme of \cite{tandon2018pir} can be impractical for a large number of databases. 
This achievable scheme was generalized to the decentralized storage placement in \cite{wei2018capacity}. In addition, linear coded storage placement at the databases has been analyzed in \cite{Banawan2018coded} and \cite{banawan2019improved}. Furthermore, Tian {\em et al.} \cite{tian2018PIR} used Shannon theoretic approach to analyze the SC-PIR problem for the canonical case of $K=2$ and $N=2$ and proposed the optimal linear scheme. More interestingly, they also showed that non-linear scheme can use less storage than the optimal linear scheme.

The SC-PIR problem was firstly generalized by Banawan {\em et al.} in \cite{banawan2019capacity} to study the case where DBs have heterogeneous storage requirements. In this setting, the storage capacity of the $N$ databases are defined by a vector\footnote{$\mathbb{R}^N_+$ denotes the set of non-negative real-valued vectors in $N$-dimensional space} $\boldsymbol{\mu}\in \mathbb{R}^N_+$, such that DB$_n$ can only store up to $\mu[n]KL$ bits and $0~\leq~\mu[n]~\leq~1$. Surprisingly, 
the authors in \cite{banawan2019capacity} showed that the capacity of heterogeneous SC-PIR is the same as homogeneous SC-PIR where $t~=~\sum_{n=1}^{N}\mu[n]$. Furthermore, the authors translated the storage placement problem into a linear program (LP). A relaxed version of the LP demonstrated that, to achieve capacity, sub-message sets should be stored at $t$ DBs (or $\lfloor t \rfloor$ and $\lceil t \rceil$ DBs for non-integer $t$). The authors also showed the existence of a solution to the LP for general $N$. However, an explicit placement solution was only derived for $N=3$ DBs. For general $N$, the LP has $O(\exp N)$ variables, representing the potential sub-messages. Hence, 
this scheme has a high complexity for large $N$.

In this paper, we propose capacity-achieving SC-PIR schemes which require a exponentially less 
number of sub-messages per message in terms of $N$ compared to the schemes proposed in \cite{tandon2018pir,wei2018capacity,attia2018capacity,banawan2019capacity}. 
We use a design framework which utilizes previously developed FS-PIR schemes for delivery and design new storage placement schemes. 
Moreover, our storage placement requires at most $N$ sub-messages per message when $t$ is an integer.\footnote{This does not include the number of sub-messages necessary for query generation. By using the query generation technique of \cite{tian2018capacity}, the total number of sub-messages to achieve heterogeneous SC-PIR capacity is $N\times(N-1)$.} For homogeneous SC-PIR, we abandon the idea of using classical shared link coded caching approaches and propose two novel combinatorial schemes.
Based on the sufficient conditions to achieve capacity in heterogeneous SC-PIR problem shown in \cite{banawan2019capacity}, we show that the storage placement problem can be translated to a {\em filling problem (FP)}. 
Instead of deriving an explicit LP solution, alternatively, we approach the problem by proposing an iterative algorithm which places a sub-message set at $t$ DBs in each iteration when $t$ is an integer. 
Finally, while our proposed SC-PIR schemes only operate on integer $t$, we derive a method to convert a non-integer $t$ storage placement problem into two integer $t$ storage placement problems.

\paragraph*{Our Contributions in this paper are as follows}
\begin{enumerate}
\item We provide a general design methodology for the SC-PIR problem by establishing a generic connection between the FS-PIR and SC-PIR problems. Based on this connection, a SC-PIR scheme can be readily designed from any given FS-PIR scheme.
\item We propose two storage placement schemes for homogeneous SC-PIR which require at most $N$ sub-messages per message without the consideration of the number of sub-messages necessary for query generation.
\item We propose an iterative storage placement algorithm which solves the heterogeneous SC-PIR placement problem for general $N$ and integer $t$ which requires at most $N$ iterations and $N$ sub-messages per message.
\item We expand our results to allow for non-integer $t$.
\end{enumerate}

The remainder of this paper is organized as follows. In Section \ref{sec: problem}, we describe the problem formulation of SC-PIR. In Section \ref{sec: des_arch}, we present a design architecture for SC-PIR storage placement schemes. The homogeneous SC-PIR storage designs are presented in Section \ref{sec: homo}. In Section \ref{sec: het}, we use the sufficient conditions of SC-PIR capacity to translate the heterogeneous SC-PIR storage placement problem into an equivalent filling problem. In Section \ref{sec: iter_des}, we develop an iterative solution to the filling problem and analyze its convergence. In Section \ref{sec: nonintt}, we expand our designs for non-integer $t$. In Section \ref{sec: Discussion} we discuss this work and future directions. Concluding remarks are given in Section \ref{sec: Conclusion}.

\paragraph*{Notation Convention}
We use $|\cdot|$ to represent the cardinality of a set or the length of a vector. 
Also $[n] := 1,2,\ldots,n$ and $[n_1:n_2] = n_1,n_1+1,\ldots ,n_2$. A bold symbol such as $\boldsymbol{a}$ indicates a vector and $a[i]$ denotes the $i$-th element of $\boldsymbol{a}$. $\mathbb{R}^n_+$ is the set of non-negative reals in $n$-dimensional space and $\mathbb{Z}^+$ is the set of all positive integers. $\Delta_n \subset \mathbb{R}^n_+$ is the unit simplex, which represents the set of all vectors with $n$ non-negative elements that sum to $1$.

\section{Problem Formulation}
\label{sec: problem}

There are $K$ independent messages, $W_1 ,\ldots , W_K$, each of size $L$ bits.
\begin{align}
H(W_1, \ldots , W_K) &= H(W_1) + \cdots + H(W_K) \\
H(W_1) &= \cdots = H(W_K) = L.
\end{align}
The messages are collectively stored in an uncoded fashion among $N$ non-colluding DBs, labeled as DB$_1,\ldots,$ DB$_N$. The storage capacity of the DBs are defined by a vector $\boldsymbol{\mu} \in\mathbb{R}^N_+$ where, for all $n\in [N]$, DB$_n$ has the storage capacity of $\mu[n] K L$ bits and $0 < \mu[n] \leq 1$. Furthermore, for all $n\in [N]$,  define $Z_n$ 
as the storage contents of DB$_n$ such that
\begin{align}
\forall n\in[N], \;\; H(Z_n) \leq \mu [n]KL.
\end{align}
Also, we define
$t \triangleq \sum_{n=1}^{N}\mu[n]$
as the number of times each bit of the messages is stored among the DBs. To design an achievable PIR scheme we assume $t \geq 1$ so that each bit of the messages can be stored at least once across the DBs.
A user makes a request $W_k$ and sends a query $Q_n^{[k]}$, which is independent of the messages, to each DB $n\in [N]$,
\begin{align}
\forall k \in [K], \;\; I(W_1,\ldots, W_K ; Q_1^{[k]} , \ldots , Q_N^{[k]}) = 0.
\end{align}
Each BD $n\in [N]$ sends an answer $A_n^{[k]}$ such that
\be
\label{eq: PIR 1}
\forall k \in [K], \;\; \forall n \in [N], \;\; H(A_n^{[k]} | Z_n , Q_n^{[k]}) = 0.
\ee
Furthermore, given the answers from all the databases, the user must be able to recover the requested message and therefore,\footnote{In this work, we explore zero-error PIR schemes.}
\be
\label{eq: PIR 2}
H(W_k | A_1^{[k]}, \ldots , A_n^{[k]} , Q_1^{[k]} , \ldots , Q_n^{[k]}) = 0.
\ee

The user generates queries in a manner to ensure privacy such that no DB can infer which message the user desires, {\em i.e.} for all $n\in [N]$ 
\be
\label{eq: PIR 3}
I(k; Q_n^{[k]}, A_n^{[k]}, W_1, \ldots, W_K, Z_1, \ldots, Z_N) = 0.
\ee
Let $D$ be the total number of downloaded bits
\begin{align}
D = \sum_{n=1}^{N}H\left(A_n^{[k]}\right).
\end{align}
Given  $\boldsymbol{\mu}$, we say that a pair $(D,L)$ is achievable if there exists a SC-PIR scheme with rate
\begin{align}
R \triangleq \frac{L}{D}
\end{align}
 that satisfies  (\ref{eq: PIR 1})-(\ref{eq: PIR 3}). The SC-PIR capacity is defined as
\be
C^*(\boldsymbol{\mu}) = \sup\{R: (D,L) \text{ is achievable}\}.
\ee

\section{SC-PIR Design Architecture and Achievable Rate}
\label{sec: des_arch}

In this section, we provide a general architecture for SC-PIR scheme design. We split the SC-PIR problem into placement and delivery phases and then propose a particular placement and choice of delivery schemes. The achievable rate of the proposed design is established.

\textit{Placement}: Define a vector $ \boldsymbol{\alpha} = [ \alpha_1 , \ldots , \alpha_F]$, 
where $F \in \mathbb{Z}^+$, $\sum_{i=1}^{F} \alpha_i =1$, and $\alpha_f ,\forall f \in [F]$ is rational number such that $\alpha_f L \in \mathbb{Z}^+$.
For all $k \in [K]$, we divide message $W_k$ into $F$ disjoint sub-messages $W_k =  W_{k,1} , \ldots ,  W_{k,F}$ such that for all $f \in [F]$, $|W_{k,f}| = \alpha_f L$ bits.  
For all $f \in [F]$, let
\be
\mathcal{M}_f \triangleq \bigcup\limits_{k \in [K]} W_{k,f},
\label{eq_M_f}
\ee
and $\mathcal{N}_f \subseteq [N]$ be a non-empty subset of DBs 
which have the sub-messages in $\mathcal{M}_f$ locally available to them. The storage contents of database $n \in [N]$ is
\be
Z_n = \left \{ \mathcal{M}_f : f \in [F], n \in \mathcal{N}_f \right \},
\label{eq_Z_n}
\ee
where we have the requirement that for any $n \in [N]$,
\be
\sum_{\left \{ f : f \in [F], n \in \mathcal{N}_f\right \} } \alpha_f \leq \mu[n].
\label{eq_mu_F}
\ee

\textit{Delivery}: Given that a user requests file $W_\theta$ for some $\theta \in [K]$, we do the following. For all $f\in [F]$, using a FS-PIR scheme, the user generates a query to privately download $W_{\theta , f}$ from the databases in $\mathcal{N}_f$. In other words, a SC-PIR scheme can be found by applying a FS-PIR scheme to each set of databases $\mathcal{N}_f$. Changing the choice of the FS-PIR scheme or the definitions of $\Nc_f$ will result in new SC-PIR schemes.

In Appendix \ref{sec: priv_pf}, we prove that this approach is private.
The rate of the SC-PIR scheme, as a function of storage placement and rate of the implemented FS-PIR schemes, is given in the following theorem.

\begin{theorem}
\label{th: full2constr}
Given $N,K,F \in \mathbb{Z}^+$ and $\boldsymbol{\alpha}$, 
split each of the $L$-bit messages $W_1 , \ldots , W_K$  into $F$ sub-messages of size $\alpha_1 L , \ldots , \alpha_F L$ and store them at sets of databases $\mathcal{N}_1 , \ldots , \mathcal{N}_F \subseteq [N]$, respectively.  Given a set of FS-PIR schemes with achievable rates $R_1 , \ldots , R_F$, the achievable rate of privately downloading $W_\theta$, $\theta \in [K]$, from the $N$ storage constrained databases is
  \be
  R = \left(\frac{\alpha_1}{ R_1} + \frac{\alpha_2}{ R_2} + \cdots + \frac{\alpha_F}{ R_F} \right)^{-1}. \label{eq: con_st_rate}
  \ee
\end{theorem}

\begin{IEEEproof}
We count the number of downloaded bits. For all $f \in [F]$,
$R_f = \frac{\alpha_f L}{D_f}$
where $D_f$ is the number of downloaded bits necessary to privately download $W_{\theta , f}$ of size $\alpha_f L$ bits from the databases in $\mathcal{N}_f$. Therefore, the total number of bits required to privately download the entirety of $W_\theta$ is
\begin{align}
D &= D_1 + D_2 + \cdots + D_F \nonumber
= L \left( \frac{\alpha_1  }{ R_1} +\frac{\alpha_2 }{ R_2} + \cdots + \frac{\alpha_F }{ R_F}\right).
\end{align}
Since $R = \frac{L}{D}$, we obtain  (\ref{eq: con_st_rate}).
\end{IEEEproof}

\section{Homogeneous SC-PIR Placement Schemes}
\label{sec: homo}
In this section, we present two placement designs which achieve capacity for the homogeneous SC-PIR problem adopting capacity achieving FS-PIR schemes for delivery. We begin with an example that leads to the first placement design which requires a constant $F=\frac{N}{t}=\frac{1}{\mu_0}$ sub-messages per message, where $\mu_0 = \mu[1] = \cdots = \mu [N]$. This design requires that $\frac{N}{t}\in\mathbb{Z}^+$. In the second design, we have $F=N$, but allow $\frac{N}{t}\notin\mathbb{Z}^+$. Both schemes have the additional requirement that $t\in \mathbb{Z}^+$. This restriction will be removed in Section \ref{sec: nonintt} where we extend our designs to $t\notin\mathbb{Z}^+$.

\subsection{A Homogeneous SC-PIR Example}
\label{sec: ex1}
In this section, we provide an example of a homogeneous SC-PIR solution which does not use coded caching storage placement designs. Surprisingly, we show that coded caching placement is not necessary to achieve SC-PIR capacity, and we significantly reduce the number of sub-messages compared to the state-of-the-art SC-PIR scheme of \cite{attia2018capacity}.

Consider $N=4$ DBs labeled as DB$1$ through DB$4$. Collectively, the DBs store $K=3$ messages, denoted by $A$, $B$ and $C$. Each message is comprised of $L=16$ bits. Moreover, each DB has the storage capacity of up to $24$ bits, or half of all $3$ messages, and therefore, $\mu [1] = \cdots = \mu [4]= \frac{1}{2}$.

\textit{Placement}:
To define the placement, we split each message as follows.
\begin{align}
A &= \left\{ a_i^j : i \in [2], j \in [8] \right\} \\
B &= \left\{ b_i^j : i \in [2], j \in [8] \right\} \\
C &= \left\{ c_i^j : i \in [2], j \in [8] \right\}.
\end{align}
Then, the storage contents of the DBs are defined to be
\begin{align}
Z_1 = Z_2 &= \left\{ a_1^j : j\in [8] \right\} \cup \left\{ b_1^j : j\in [8] \right\} \cup\left\{ c_1^j : j\in [8] \right\} \\
Z_3 = Z_4 &= \left\{ a_2^j : j\in [8] \right\} \cup \left\{ b_2^j : j\in [8] \right\} \cup\left\{ c_2^j : j\in [8] \right\}.
\end{align}
Each database stores $8$ out of $16$ bits of each message. Databases $1$ and $2$ have the same storage contents, but do not have any storage contents in common with databases $3$ and $4$. Likewise, databases $3$ and $4$ have the same storage contents.

\textit{Delivery}:
The placement has reduced the SC-PIR problem into two independent FS-PIR problems; one consists of DB $1$ and $2$, and the other consists of DB $3$ and $4$. Subsequently, we can adopt the achievable FS-PIR scheme of \cite{sun2017capacity} to generate the queries for each pair of DBs separately. The queries of a user that desires message A are shown in Table \ref{table: ex_1}.

\begin{table}[h!]
\vspace{-0.4cm}
\normalsize
\centering
{\small
\caption{ Storage Constrained PIR, $N=4$, $K=3$, $\mu = \frac{1}{2}$}
\vspace{-0.2cm}
 \label{table: ex_1}
\begin{tabular}{ |>{\centering}m{1.7cm}|>{\centering}m{1.7cm}||>{\centering}m{1.7cm}|>{\centering}m{1.7cm}| }
 \hline
 DB$1$ & DB$2$ & DB$3$ & DB$4$ \\
 \hline
 $a_1^5 \quad b_1^8 \quad c_1^6$ & $a_1^1 \quad b_1^3 \quad c_1^1$ & $a_2^5 \quad b_2^7 \quad c_2^4$ & $a_2^2 \quad b_2^6 \quad c_2^2$ \\
 \hline
 $a_1^6+b_1^3$ & $a_1^3+b_1^8$ & $a_2^1+b_2^6$ & $a_2^7+b_2^7$ \\
 $a_1^7+c_1^1$ & $a_1^8+c_1^6$ & $a_2^6+c_2^2$ & $a_2^8+c_2^4$ \\
 $b_1^6+c_1^5$ & $b_1^7+c_1^3$ & $b_2^3+c_2^6$ & $b_2^8+c_2^7$ \\
 \hline
 $a_1^2+b_1^7+c_1^3$ & $a_1^4+b_1^6+c_1^5$ & $a_2^3+b_2^8+c_2^7$ & $a_2^4+b_2^3+c_2^6$ \\
 \hline
\end{tabular}
}
 \end{table}

Note that, for ease of disposition, we have chosen to use the FS-PIR scheme of \cite{sun2017capacity}. In fact, any FS-PIR scheme can be used. For example, the FS-PIR of \cite{tian2018capacity} requires only $t-1 = 1$ sub-messages per message for delivery, as opposed to $t^K = 8$ sub-messages for \cite{sun2017capacity}. If the scheme of \cite{tian2018capacity} was used, the size of each message could be $2$ bits, instead of $16$.

The total number of downloaded bits is $D = 28$.   Thus, for this scheme {$R=\frac{L}{D}=\frac{16}{28}=\frac{4}{7}$}, which achieves the capacity 
of $(1+\frac{1}{t}+\frac{1}{t^2})^{-1}=(1+\frac{1}{2}+\frac{1}{2^2})^{-1}=\frac{4}{7}$. Compared to the SC-PIR scheme of \cite{attia2018capacity} which requires $L = {N \choose t}t^K={4 \choose 2}2^3=48$ bits, the proposed SC-PIR requires only $L=16$ bits.


Privacy is ensured since the FS-PIR scheme of \cite{sun2017capacity} is used to privately download half of message $A$ from DB$1$ and DB$2$ and the other half from DB$3$ and DB$4$. The query to each database is symmetric such that for each bit of $A$ that is requested, a bit each from $B$ and $C$ are also requested. All coded pairs of bits from the $3$ messages are requested an equal number of times.
Ultimately, the user can decode all  bits of message $A$, because downloaded bits of $B$ and $C$ can be used for decoding (see Table~\ref{table: ex_1}).
In the following, we generalize this example.

%

\subsection{General SC-PIR Scheme when $\frac{N}{t} \in \mathbb{Z}^+$}
\label{subsec_SC-PIR1}

\subsubsection{Storage Placement Scheme}
Given $N \in \mathbb{Z}^+$ and $t \in [N]$ such that $\frac{N}{t} \in \mathbb{Z}^+$, let $F = \frac{N}{t}$ and for each $k \in [K]$, split message $W_k$ into $\frac{N}{t}$ disjoint, equal-size sub-messages, $W_{k,1} , \ldots , W_{k,\frac{N}{t}}$. Furthermore, split the $N$ databases into $\frac{N}{t}$ disjoint groups of size $t$ labeled as $\mathcal{N}_1 , \ldots , \mathcal{N}_{\frac{N}{t}}$. For each $f \in \left[ \frac{N}{t} \right]$, the sub-messages of
\be
M_f = \bigcup\limits_{k\in[K]} W_{k,f}
\ee
are stored at every database of $\mathcal{N}_f$.

\subsubsection{PIR Scheme}
\label{sec: pirsc1}
A user desires to privately download message $W_\theta$ for some $\theta \in [K]$. For each $f \in \left[ \frac{N}{t} \right]$, the user generates a query using a capacity achieving FS-PIR scheme to privately download $W_{\theta , f}$ from the $t$ databases in $\mathcal{N}_f$. The user combines the downloaded sub-messages, $W_{\theta , 1} , \ldots , W_{\theta , \frac{N}{t}}$ to recover the desired message $W_{\theta}$.

To implement this SC-PIR scheme, each message is split into $\frac{N}{t}$ equal-size, disjoint sub-messages for storage placement. Furthermore, the adaptation of the FS-PIR scheme requires that each sub-message is further split. For example, by using the scheme of \cite{tian2018capacity}, the resulting SC-PIR requires a minimum message size of $L = \frac{N}{t} \cdot (t-1)=   N - \frac{\mu_0}{N}$ bits.

\subsubsection{Achievable Rate}
The achievable rate of this scheme is summarized as follows.

\begin{corollary}
\label{co: new_cons1}
   Given $N,K,$ and $\mu_0 \in \left[\frac{1}{N},1\right]$, such that $t = \mu_0 N \in [N]$ and $\frac{N}{t} \in \mathbb{Z}^+$, for a user to privately download one of $K$ $L$-bit messages from $N$ databases with a storage capacity of $\mu_ K L$ bits, the achievable rate is
  \be
  R = \left(1 + \frac{1}{t} + \frac{1}{t^2} + \cdots +\frac{1}{t^{K-1}} \right)^{-1}. \label{eq: sc1_rate_1}
  \ee
\hfill$\square$
\end{corollary}

It was shown in \cite{attia2018capacity} that (\ref{eq: sc1_rate_1}) is the capacity of SC-PIR for $t\in \mathbb{Z}^+$. While we do not directly prove Corollary~\ref{co: new_cons1} here, in Section \ref{sec: suff_cond} we present a set of sufficient conditions, which this scheme satisfies, for a SC-PIR scheme to meet the capacity.


\subsection{SC-PIR Scheme when $\frac{N}{t} \notin \mathbb{Z}^+$}
\label{sec-new-SC-PIR}
In the following we present a SC-PIR scheme which allows $\frac{N}{t} \notin \mathbb{Z}^+$ and requires $F=N$ number of sub-messages per message for the placement phase. We first provide an example, then discuss the general scheme.

\subsubsection{A SC-PIR Example when $\frac{N}{t} \notin \mathbb{Z}^+$}

In this example, $N=5$ DBs, labeled DB$1$ through DB$5$, collectively store $K=2$ messages, $A$ and $B$, and each has a size of $L=15$ bits. Each DB stores an $\mu_0 = \frac{3}{5}$ fraction of the $2$-message library ($t=\mu_0 N=3$).

\textit{Placement}: Each message is split as follows.
\begin{align}
  A &= \left\{ a_i^j : i \in [5], j \in [3] \right\}, \quad
  B = \left\{ b_i^j : i \in [5], j \in [3] \right\}.
\end{align}
By this labeling, we have split the messages in two phases. The first splitting phase, denoted by the subscript, define the placement and determines which DBs store these bits. The second splitting, denoted by the superscript, is necessary to perform the FS-PIR scheme. For all $f \in [5]$, define
\be
M_f = \bigcup\limits_{j \in [3]}\left( a_f^j \cup b_f^j \right)
\ee
and let the set of databases $\mathcal{N}_f = [-2:0]\oplus_N f$ locally store the bits of $M_f$.\footnote{We impose the following notation: $a \oplus_N b = (a+b-1 \mod N) +1$ and $[a_1:a_2] \oplus_N b =\left\{ a' \oplus_N b : a' \in [a_1 : a_2 ]\right \}$.}
 Note that as opposed to the SC-PIR scheme described in Section \ref{sec: ex1} where the sets of databases $\{\mathcal{N}_f, f=1,\cdots, F\}$ are mutually exclusive, here we allow them to overlap and hence removing the integer constraint of $\frac{N}{t} \in  \mathbb{Z}^+ $.
As a result, the bits of message $A$ stored at DB $n\in[5]$ are
\be
Z_n = \bigcup\limits_{f \in \left\{ [0:2] \oplus_N n \right \}}M_f.
\ee
For instance, DB$2$ stores all bits $a_i^j$ and $b_i^j$ such that $i \in [2:4] $ and DB$5$ stores all bits $a_i^j$ and $b_i^j$ such that $i \in \left\{ 5,1,2 \right\} $.

\begin{table}[h!]
\vspace{-0.4cm}
\normalsize
\centering
{\small
\caption{ Storage Constrained PIR, $N=5$, $K=2$, $\mu = \frac{3}{5}$}
\vspace{-0.2cm}
\label{table: 2}
\begin{tabular}{ |>{\centering}m{1.32cm}|>{\centering}m{1.32cm}|>{\centering}m{1.32cm}|>{\centering}m{1.32cm}|>{\centering}m{1.32cm}| }
 \hline
 DB$1$ & DB$2$ & DB$3$ & DB$4$ & DB$5$ \\
 $( 1,2,3)$ &$( 2,3,4)$ &$( 3,4,5)$ &$( 4,5,1)$ &$( 5,1,2)$ \\
 \hline
 ${\color{red} a_1^3} \quad {\color{red}b_1^2} $ & $a_2^3 \quad b_2^2$ & $a_3^1 \quad b_3^3 $ & $a_4^2 \quad b_4^3 $ & $a_5^2 \quad b_5^1$ \\
 \hline
 $a_2^1+b_2^2$ & $a_3^3+b_3^3$ & $a_4^3+b_4^3$ & $a_5^1+b_5^1$ & ${\color{red} a_1^2+b_1^2}$ \\
 $a_3^2+b_3^3$ & $a_4^1+b_4^3$ & $a_5^3+b_5^1$ & ${\color{red}a_1^1+b_1^2}$ & $a_2^2+b_2^2$ \\
 \hline
\end{tabular}
}
\end{table}
\textit{Delivery}: The queries of a user that desires to privately download message $A$ are shown in Table \ref{table: 2}. The top row of the table  contains  database labels and the $3$-tuple below each database label defines the subscripts of the bits that are locally available to that database. The remaining three rows of the table  show the queries of the user. 
The user adopts the FS-PIR scheme of \cite{sun2017optimal} to design queries. For instance, to obtain bits $\{a_1^j, j\in[3]\}$, the user applies the FS-PIR to DB$1$, DB$4$, and DB$5$. In the first round, the user obtains $a_1^3$ from DB$1$. In the second round, the user can decode $a_1^1$ from DB$4$'s transmission of $a_1^1 + b_1^2$ because the user had already received $b_1^2$ from the first round transmission of DB$1$ in round $1$. Similarly, the user decodes $a_1^2$ from DB$5$'s transmission of $a_1^2 + b_1^2$.  These transmissions are highlighted in red in Table \ref{table: 2}.
To ensure privacy, the queries are symmetric and no bit is requested more than once from any one database. In this example, $D=20$ bits are downloaded and the rate is $R = \frac{3}{4}$. Comparing to the state-of-the-art SC-PIR scheme of \cite{attia2018capacity}, the rate is the same, but $L$ has been reduced from ${N \choose t}t^K = {5 \choose 3}3^2=90$ to $Nt^{K-1} = 5 \cdot 3^{2-1}= 15$.

Similar to the previous example, we use the FS-PIR scheme of  \cite{sun2017optimal} for ease of disposition. The size of each message could be reduced to $L=N(t-1)=10$ by using the FS-PIR scheme of \cite{tian2018capacity}.

\subsubsection{Storage Placement Scheme}
For each $k \in [K]$, message $W_k$ is split into $F=N$ disjoint equal-size sub-messages $W_{k,1}, \ldots , W_{k,N}$. For all $f\in [N]$, define a set of sub-messages
$M_f = \cup_{k \in [K]}W_{k,f}$
which is locally stored at the set of databases $\mathcal{N}_f = [-(t-1):0]\oplus_N f$.

\subsubsection{PIR Scheme}
The PIR scheme is the same as the previous scheme in Section \ref{sec: pirsc1}: a user who desires to download $W_\theta$ uses a FS-PIR scheme to privately download $W_{\theta,f}$ from $\mathcal{N}_f$ for all $f\in [F]$.


\subsubsection{Achievable Rate}
The achievable rate of this SC-PIR scheme is summarized in the following corollary.

\begin{corollary}
\label{co: new_cons2}
 Given $N,K,$ and $\mu \in \left[\frac{1}{N},1\right]$, such that $t = \mu N \in [N]$ and $L =  N t^{K-1}$, for a user to privately download one of $K$ $L$-bit messages from $N$ databases, each with a storage capacity of $\mu K L$ bits, the rate is
\be
R = \left(1 + \frac{1}{t} + \frac{1}{t^2} + \cdots +\frac{1}{t^{K-1}} \right)^{-1}. \label{eq: sc1_rate}
\ee
\end{corollary}

The results of Section \ref{sec: suff_cond} demonstrate that this SC-PIR scheme satisfies the sufficient conditions to meet the capacity.  This proves Corollary \ref{co: new_cons2}.

\section{Translating Heterogeneous SC-PIR to a Filling Problem}
\label{sec: het}
In this section, we translate the heterogeneous PIR storage placement problem into an equivalent filling problem (FP). We start by stating a set of sufficient conditions to meet SC-PIR capacity. Then, we present an example of an FP solution to achieve capacity in an equivalent heterogeneous SC-PIR storage placement problem.
Next, we formally define the FP and explain its connection to the sufficient conditions of Lemma \ref{th: suff_cond}. Moreover, we define a set of conditions which guarantees a FP solution and demonstrate that there always exists a heterogeneous SC-PIR storage placement solution when $t\in\mathbb{Z}^+$.
This section motivates the remainder of this paper which aims to find a solution to the heterogeneous SC-PIR placement problem by solving an equivalent FP.

\subsection{Sufficient Conditions to Achieve Capacity for SC-PIR}
\label{sec: suff_cond}


In Lemma \ref{th: suff_cond}, we provide sufficient conditions for a storage placement scheme to achieve the SC-PIR capacity. These conditions are proved in \cite{woolsey2019newPIR} for the homogeneous case and \cite{banawan2019capacity} for the more general heterogeneous case. We refer the readers to \cite{woolsey2019newPIR} and \cite{banawan2019capacity} for the proof of Lemma \ref{th: suff_cond}.

\begin{lemma}
\label{th: suff_cond}
Given $N,K,F \in \mathbb{Z}^+$ and $\boldsymbol{\alpha}$, 
split each of the $L$-bit messages $W_1 , \ldots , W_K$  into $F$ sub-messages of size $\alpha_1 L , \ldots , \alpha_F L$ and store them at sets of databases $\mathcal{N}_1 , \ldots , \mathcal{N}_F \subseteq [N]$ according to equations (\ref{eq_M_f})-(\ref{eq_mu_F}). For all $n\in[N]$, DB$_n$ has a storage capacity of $\mu[n] KL$ bits, $0 \leq \mu[n] \leq 1$ and let $t = \sum_{n=1}^{N}\mu[n]$.
Assume that a user requests file $W_\theta$ for some $\theta \in [K]$. A SC-PIR scheme is obtained if for all $f\in [F]$, the user generates a query to privately download $W_{\theta , f}$ from the databases in $\mathcal{N}_f$ using a capacity-achieving FS-PIR scheme.
The resulting SC-PIR scheme is capacity-achieving if the storage placement satisfies one of the following two conditions:
  \begin{enumerate}[(a)]
    \item if $t\in\mathbb{Z}^+$, then $|\mathcal{N}_f| = t$, $\;\forall f \in [F]$.
    \item if $t \notin \mathbb{Z}^+$, then $|\mathcal{N}_f| \in \{ \lfloor t \rfloor , \lceil t \rceil \}$, $\;\forall f \in [F]$,
    \be
    \label{eq: notintt1}
    \sum_{f:|\mathcal{N}_f| = \lfloor t \rfloor}\alpha_f=\lceil t \rceil - t
    \ee
     and
     \be
     \label{eq: notintt2}
    \sum_{f:|\mathcal{N}_f| = \lceil t \rceil}\alpha_f = t - \lfloor t \rfloor.
    \ee
  \end{enumerate}
\end{lemma}

In the following example, $t\in \mathbb{Z}^+$ and we demonstrate how to iteratively fill the DB storage where each iteration fills some contents of $t$ DBs.

\begin{figure*}
\centering
\centering \includegraphics[width=16.25cm]{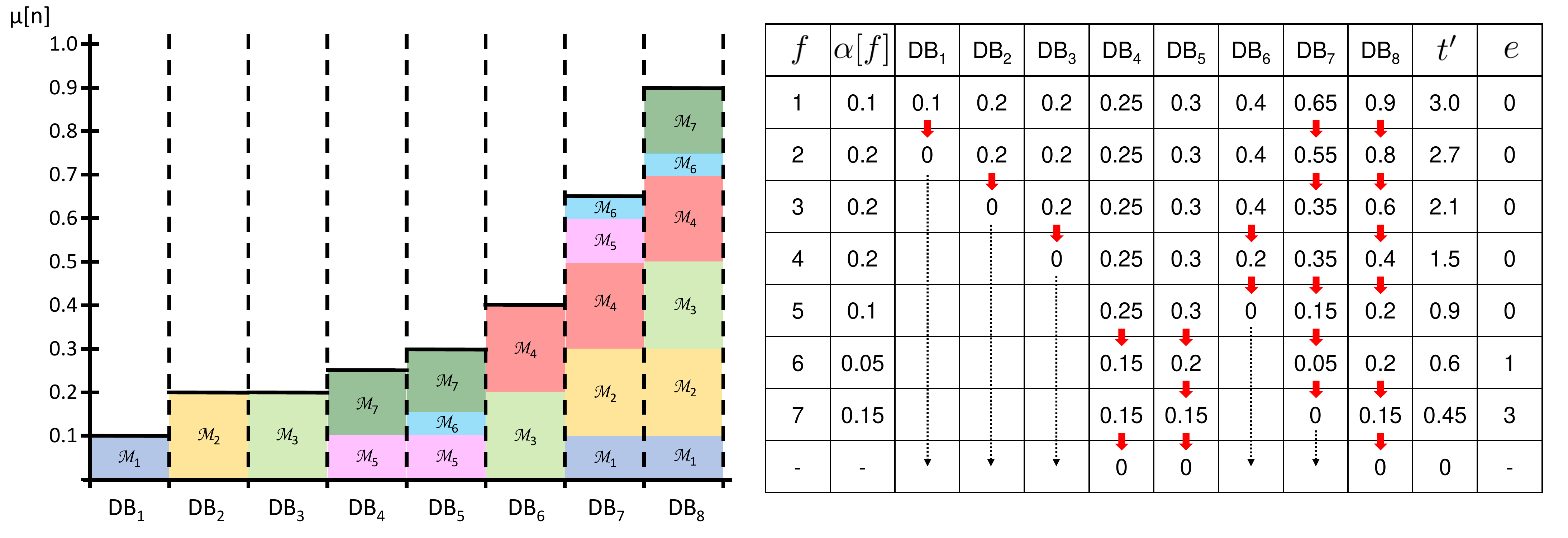} 
\vspace{-0.4cm}
\caption{~\small A solution to the filling problem using Algorithm \ref{algorithm:1} when $t=3$ and $\boldsymbol{\mu}= [0.1,\; 0.2,\; 0.2,\; 0.25,\; 0.3,\; 0.4,\; 0.65,\; 0.9]$. (left) A bar graph depicting the storage requirements of the DBs and the storage placement solution. (right) A table representing the remaining storage of the DBs for each iteration. The red arrows highlight which DBs are assigned a sub-message subset in each iteration. 
}
\label{fig: exp1}
\vspace{-0.4cm}
\end{figure*}

\subsection{A Heterogeneous SC-PIR Example}

Let $N=8$ and the storage constraints of the DBs are
\be
\label{eq: mu_ex}
\boldsymbol{\mu} = [0.1,\;\; 0.2,\;\; 0.2,\;\; 0.25,\;\; 0.3,\;\; 0.4,\;\; 0.65,\;\; 0.9].
\ee
For example, by this notation, DB$_6$ has a storage capacity of $\frac{4}{10}KL$ bits. By summing the elements of $\boldsymbol{\mu}$, we obtain   $t=3$.

To define the storage placement, the $K$ messages are divided into $F$ disjoint sub-message sets, $\mathcal{M}_1,\ldots,\mathcal{M}_F$, such that each sub-message set contains a sub-message of equal size from each of the $K$ messages. Then, each sub-message set, $\mathcal{M}_f$, is stored at some subset of DBs $\mathcal{N}_f\subseteq[N]$. 
In \cite{banawan2019capacity}, it was proposed to solve a LP to determine these sub-messages and DB sets to achieve the heterogeneous SC-PIR capacity. However, the LP has an exponential number of variables with respect to $N$ such that it may not be practical for large $N$. 
Since the capacity can be achieved if each sub-message set, $\mathcal{M}_f$, is stored at exactly $t=3$ DBs, we realize that this translates to a ``filling problem" (FP) where our goal is to iteratively fill the storage of the DBs and in each iteration we fill some available storage in exactly $3$ DBs.

We propose an iterative scheme to solve this filling problem where each iteration aims to fill  the DB with the least remaining storage. In the first iteration, we define a sub-message set, $\mathcal{M}_1$, which contains $\mu[1]L=\frac{1}{10}L$ arbitrary bits from each of the $K$ messages and assign $\mathcal{M}_1$ to the DB subset $\mathcal{N}_1=\{1,7,8 \}$. Notice that $\mathcal{M}_1$ contains $\mu[1]KL$ bits and there is no remaining available storage at DB$_1$ after this iteration. 
After this iteration, the question arises whether or not this iteration yields a valid placement (for future iterations). Later in Section \ref{sec: exist} we define a set of necessary and sufficient conditions to determine whether a particular iteration is valid.

Next, we aim to fill the storage contents of DB$_2$ and 
let $\mathcal{M}_2$ 
contain $\frac{1}{5}L$ arbitrarily unpicked bits (i.e., bits are not in $\mathcal{M}_1$) from each of the $K$ messages. 
Then, $\mathcal{M}_2$ is stored at the DB subset $\mathcal{N}_2=\{2,7,8 \}$. In general, 
the idea to determine $\Nc_f$ is to choose the DB with the smallest remaining storage and the $t-1$ DBs with the most remaining available storage. This process is continued until the $5$th iteration, where filling  DB$_7$ (which has  the smallest remaining storage) would cause for an invalid filling solution. Later in Section \ref{sec: iter_des}, we discuss how to handle this by not completing filling the DB with the smallest remaining storage.

The final results of the storage placement by our newly proposed algorithm are shown in Fig. \ref{fig: exp1}. In total, there are $F=7$ sub-message sets, each of which contains a sub-message from each of the $K$ messages, and is stored at exactly $3$ DBs. A vector $\boldsymbol{\alpha} \in \Delta_F$ defines the fraction of the library that is stored (or filled) in each iteration. For example, $\alpha [1]=0.1$ and $\alpha [2] = 0.2$ correspond to the first two iterations described above. All of the values of $\boldsymbol{\alpha}$ are shown in the table of Fig. \ref{fig: exp1}. The corresponding DBs that store a sub-message subset in a particular iteration are highlighted by the red arrows in the table of Fig. \ref{fig: exp1}.

Given that a user desires to privately download $W_\theta$ for some $\theta\in[K]$, the user will privately download the sub-message of $W_\theta$ stored at DBs of $\mathcal{N}_f$ using one of the capacity achieving FS-PIR in \cite{sun2017capacity,tian2018capacity,sun2017optimal} for all $f\in[F]$. The rate of each download and the overall rate is equal to the rate of a capacity-achieving FS-PIR scheme that is privately downloading from $t=3$ DBs. In this case, the rate is
\be
R=\left(1 + \frac{1}{t} + \frac{1}{t^2} + \cdots + \frac{1}{t^{K-1}} \right)^{-1}
\ee
which was shown to be the capacity of heterogeneous SC-PIR in \cite{banawan2019capacity}.

Fig. \ref{fig: exp1} contains two additional parameters, $t'$ and $e$, which are discussed in greater detail later in this paper. Moreover, $t'$ is the sum of the cumulative normalized remaining storage of all DBs and $e$ is the  number of DBs that each has a remaining storage that is equal to $\frac{t'KL}{t}$ bits. These parameters are significant when deriving the necessary and sufficient conditions for a valid placement and proving the convergence rate of our proposed placement algorithm.

\label{sec: fill_prob}

\subsection{The Filling Problem}
The ($\boldsymbol{m},\tau$)-\textit{Filling Problem} (FP) is defined as follows:\\

Define a basis $\mathcal{B}$, containing the set of all $\{0,1\}$-vectors of length $N$, each of which consists of exactly $\tau$ $1$s. Given a vector $\boldsymbol{m} \in \mathbb{R}_{+}^N$, representing a normalized storage vector of $N$ DBs, find a $\tau$-\textit{fill} 
defined by s set of scalars $\{\alpha_{\boldsymbol{b}}\in\mathbb{R}_+: \boldsymbol{b}\in\mathcal{B}\}$ such that
\be
\sum_{\boldsymbol{b} \in \mathcal{B}}\alpha_{\boldsymbol{b}}\boldsymbol{b} = \boldsymbol{m}.
\ee


For the heterogeneous SC-PIR problem, with $t \in \mathbb{Z}^+$ the capacity achieving placement solution is equivalent to the $(\boldsymbol{\mu},t)$-FP. 

\subsection{Existence of the $(\boldsymbol{m},\tau)$-FP Solution}
\label{sec: exist}
We aim to find a set of necessary and sufficient conditions such that a solution to the $(\boldsymbol{m},\tau)$-FP exists. Given any $\boldsymbol{m} \in \mathbb{R}_+^N$ and $\tau \in \mathbb{Z}^+$, the existence of a $(\boldsymbol{m},\tau)$-FP solution is not guaranteed. For example, if $\boldsymbol{m} = [0.3,0.3,0.7]$ and $\tau = 2$, then a $(\boldsymbol{m},\tau)$-FP solution does not exist since $m[1] + m[2] < m[3]$. 
In regards to SC-PIR, $\boldsymbol{m}$ may represent the normalized remaining storage of $3$ DBs after some placement iterations with $t=2$. It is impossible to fill the remaining storage of two DBs at a time and completely fill DB$_3$. The following theorem states the necessary and sufficient conditions for a $(\boldsymbol{m},\tau)$-FP solution to exist.

\begin{theorem}\label{th: exist}
Given $\boldsymbol{m} \in \mathbb{R}_+^N$ and $\tau \in \mathbb{Z}^+$ an $(\boldsymbol{m},\tau)$-FP solution exists \textit{if and only if}
\be
\label{eq: condition FP}
m[n] \leq \frac{\sum_{i=1}^{N}m[i]}{\tau}
\ee
for all $n \in [N]$.
\end{theorem}

Theorem \ref{th: exist} is proven in Appendix \ref{sec: FPexistpf}.
This result and the proof of Theorem \ref{th: exist} have two important implications for heterogeneous SC-PIR. First, for any given $\boldsymbol{\mu}$, if $t$ is an integer then
\be
\mu[n]\leq 1 = \frac{\sum_{i=1}^{N}\mu[i]}{t}
\ee
for all $n\in[N]$. A $(\boldsymbol{\mu},t)$-FP solution exists and therefore a heterogeneous SC-PIR scheme exists which can achieve capacity.\footnote{The existence of a solution for a capacity achieving storage placement for heterogenous SC-PIR was also shown in the proof of Lemma $5$ of \cite{banawan2019capacity}. However, the proof assumes non-integer $t$ and uses different methods according to our understanding.} Second, while it is not clear how to determine the storage placement for a given $\boldsymbol{\mu}$, Theorem \ref{th: exist} suggests that there is an iterative process which can define the storage placement. In other words, if a sub-message set is assigned to a set of $t$ DBs, then we can determine if the remaining storage among all DBs has a FP solution. In this way, an iterative scheme can be defined that is guaranteed to move towards a final storage placement solution.

\section{Iterative Storage Placement Design}
\label{sec: iter_des}
Motivated by Theorem \ref{th: exist}, in this section, we develop an iterative storage placement scheme where in each iteration a sub-message of each of the $K$ messages is placed in a set of $t$ DBs. Each iteration of the placement scheme aims to fill the storage of the DB with the smallest remaining (non-zero) storage to make the remaining FP simpler. Specifically, we store a sub-message set at a set of $t$ DBs including the DB with the smallest remaining storage and the $t-1$ DBs with the largest remaining storage. We study the convergence of our algorithm and find at most $N$ iterations are necessary to define a capacity achieving SC-PIR storage placement.

\subsection{Iterative Placement Algorithm}
The filling scheme is outlined in Algorithm \ref{algorithm:1} and a single iteration is summarized as follows.

Let $N'$ be the number of DBs with non-zero remaining storage and $\boldsymbol{m}\in\mathbb{R}^{N}_{+}$ be the remaining storage of each DB normalized by $KL$. For ease of notation and WLOG we assume $m[1] \leq m[2] \leq \ldots \leq m[N]$ for any given iteration.\footnote{For correctness, in Algorithm \ref{algorithm:1}, $\boldsymbol{m}$ is not assumed to be in increasing order and the indices corresponding to the order are used as necessary.}

If $N' \geq t+1$, do the following. Let the DB subset, $\mathcal{N}$, of size $t$ include the DB with the smallest remaining (non-zero) storage and the $t-1$ DBs with the largest remaining storage. In other words,
\be
\mathcal{N}~=~\left\{ N-N'+1 , N-t+2, \ldots , N \right\}
\ee
where $m[N-N'+1]$ is the storage remaining at the DB with the smallest remaining, non-zero storage. A sub-message set is defined to be stored at the DBs of $\mathcal{N}$. Ideally, the number of bits in the sub-message set is size $m[N-N'+1]KL$ bits, however, it is possible that such a sub-message assignment prevents a FP solution for the remaining storage among the DBs (i.e., violate (\ref{eq: condition FP})). Therefore, define $t' = \sum_{n=1}^{N}m[n]$ and let
\be
\label{eq: alpha}
\alpha = \min\left( \frac{t'}{t} - m[N-t+1], m[N-N'+1] \right)
\ee
be the normalized size of each sub-message to be placed in this iteration (refer line $10$ in Algorithm \ref{algorithm:1}).\footnote{Here, $\alpha$ is equivalent to $\alpha_F$ in Algorithm \ref{algorithm:1}. The subscript, $F$, is used in Algorithm $1$ to count the number of and distinguish betweens iterations.} Following the method presented in Section \ref{sec: des_arch}, define a sub-message set, $\mathcal{M}$, containing $\alpha KL$ bits which have not been stored in a previous iteration and store $\mathcal{M}$ at the DBs of $\mathcal{N}$. Then, adjust $\boldsymbol{m}$ accordingly to reflect the remaining storage at each DB.

There is only one exception to this process which is the case where there are only $N'=t$ DBs with non-zero remaining storage. In this case, all of the remaining storage of these $t$ DBs are equal (can be shown using Theorem \ref{th: exist}). Furthermore, let $\alpha = m[N-N'+1]$ and a sub-message set of size $\alpha KL$ bits is stored at these $t$ DBs.


Note that, Algorithm \ref{algorithm:1} only operates when $N'\geq t$, because it is impossible for $N'<t$ since to have a valid FP solution, there must be at least $t$ DBs with non-zero remaining storage. In the Appendix \ref{sec: correct_alg}, we show that each iteration of Algorithm \ref{algorithm:1} is guaranteed to have a valid FP solution for the next iteration to demonstrate the correctness of the algorithm.

\begin{algorithm}
  \caption{Heterogeneous SC-PIR Storage Placement}
  \label{algorithm:1}
  \begin{algorithmic}[1]
  \item[ {\bf Input}: $\boldsymbol{\mu}$, $t$, $L$ and $W_1,\ldots, W_K$]
  \item $\boldsymbol{m} \leftarrow \boldsymbol{\mu}$
  \item $F \leftarrow 0$
  \While {$\boldsymbol{m} > \boldsymbol{0}$}
    \State $F \leftarrow F+1$
    \State $t' \leftarrow \sum_{n=1}^{N}m[n]$
    \State $\boldsymbol{\ell} \leftarrow$ indices of non-zero elements of $\boldsymbol{m}$ from smallest to largest
    \State $N'\leftarrow$ number of non-zero elements in $\boldsymbol{m}$
    \State $\mathcal{N}_F \leftarrow\{\ell [1], \ell [N'-t+2] , \ldots , \ell [N'] \}$
    \If {$N' \geq t+1$}
    \State $\alpha_F \leftarrow  \min \left(\frac{t'}{t} - m[\ell[N' - t + 1]], m[\ell[1]]\right)$
    \Else
    \State $\alpha_F \leftarrow  m[\ell[1]]$
    \EndIf
    \For {$n \in \mathcal{N}_F$}
    \State $m[n] \leftarrow m[n] - \alpha_F$
    \EndFor
  \EndWhile
  \For {$k = 1,\ldots,K$}
  \State Partition $W_k$ into $F$ disjoint sub-messages: $W_{k,1}, \ldots , W_{k,F}$ of size $\alpha_1L,\ldots,\alpha_FL$ bits respectively
  \For {$f = 1,\ldots,F$}
  \State Store $W_{k,f}$ at the DBs of $\mathcal{N}_f$
  \EndFor
  \EndFor
  \end{algorithmic}
\end{algorithm}

\subsection{Convergence}
\label{sec: conv}

Since in each iteration we fill a positive amount of remaining storage without violating {the existence conditions for a FP solution}, Algorithm \ref{algorithm:1} converges to a final solution where all DBs are completely filled. The question remains as to how many iterations are required for convergence. Moreover, the number of iterations is equal to the number of sub-messages per message, $F$, required for the storage placement. Surprisingly, we find that at most $N$ iterations are required to fill all DBs. The {result is} summarized in the following theorem.

\begin{theorem}
Algorithm \ref{algorithm:1} requires at most $N$ iterations to completely fill the DBs.
\end{theorem}
\begin{IEEEproof}
Throughout this proof, let $\boldsymbol{m}\in\mathbb{R}^{N}_{+}$ be the remaining storage of each DB at a given iteration normalized by $KL$ and WLOG $m[1] \leq m[2] \leq \ldots \leq m[N]$. Define $t'$ as the cumulative remaining normalized storage among the DBs
\be \label{eq: t_prime}
t' = \sum_{n=N-N'+1}^{N}m[n]
\ee
where $N'$ is the number of DBs with non-zero remaining storage. We observe the iterations of Algorithm \ref{algorithm:1} and label the outcome of each iteration as either a {\it complete fill} (CF) or {\it partial fill} (PF) defined below.
\begin{defn}
  A {\it complete fill} (CF) refers to an iteration where the remaining storage at the DB with the smallest remaining non-zero storage is completely filled.
\end{defn}
\begin{defn}
  A {\it partial fill} (PF) refers to an iteration where the remaining storage at the  DB with the smallest remaining non-zero storage is {\it not} completely filled.
\end{defn}

To obtain an upper bound on the number of iterations to fill the DBs, we count the maximum number of possible PFs and CFs. To do this we introduce a new variable, $e$, which counts the number of DBs with remaining normalized storage equal to $\frac{t'}{t}$ such that
\be \label{eq: x}
e = \sum_{n=1}^N \mathbbm{1}\left ( m[n] = \frac{t'}{t}  \right )
\ee
where $\mathbbm{1}\left( \cdot \right )$ is the indicator function.
The following lemma discusses the sufficient condition which guarantees a CF for a given iteration.

%
%


\begin{lemma}\label{lm: 2} If a given iteration satisfies $e=t-1$ and $N'~\geq~t+1$, then this iteration must be a CF, and $N'$ will be reduced by at least $1$ after that iteration.
\end{lemma}

\begin{IEEEproof}
By using the condition $e=t-1$ and (\ref{eq: t_prime}), we obtain
\be
 m[N-N'+1] + \cdots + m[N-t+1] + (t-1)\frac{t'}{t}=t'. 
 \ee
Therefore,
 \be
 m[N-N'+1] + m[N-t+1] \leq \frac{t'}{t},
\ee
and
\be
m[N-N'+1] \leq \frac{t'}{t} - m[N-t+1].
\ee
By (\ref{eq: alpha}), during this iteration, $\alpha KL$ bits are stored at DB$_{N-N'+1}$ where $\alpha = m[N-N'+1]$. This completes the proof of Lemma \ref{lm: 2}.
\end{IEEEproof}

\begin{lemma}\label{lm: 3}
If a given iteration satisfies $e\leq t-1$ and $N'~\geq~t+1$, then $e$ will not decrease after that iteration. Moreover, if the iteration is a PF then $e$ will be increased by at least $1$ after that iteration.
\end{lemma}

\begin{IEEEproof}
We prove Lemma \ref{lm: 3} as follows. The $e$ DBs with normalized remaining storage equal to $\frac{t'}{t}$ are included in the set of $t-1$ DBs with the largest remaining storage since $m[n]\leq \frac{t'}{t}$ for all $n\in[N]$. Therefore, after an iteration, the normalized remaining storage of these $e$ DBs are reduced by $\alpha$ and their normalized remaining storage becomes $\frac{t'}{t}-\alpha$. Furthermore, let $t''$ be the sum of normalized storage after this iteration. Moreover,
\be \label{eq: t_pp}
t'' = t' - t\alpha
\ee
and $\frac{t''}{t} = \frac{t'}{t}-\alpha$. Hence, whether the iteration is a PF or CF, $e$ is not decreasing from one iteration to the next.

Next, consider the case where the iteration is a PF, then by (\ref{eq: alpha}), we obtain $m[N-N'+1] > \frac{t'}{t}-m[N-t+1]$. Therefore, $\alpha = \frac{t'}{t}-m[N-t+1]$. Furthermore, by (\ref{eq: t_pp}), $\frac{t''}{t} = m[N-t+1]$. As the normalized remaining storage at DB$_{N-t+1}$ remains $m[N-t+1]=\frac{t''}{t}$ and this DB is not included in the $e$ DBs with $\frac{t'}{t}$ normalized remaining storage,\footnote{This is because that if this DB is included the $e$ DBs with $\frac{t'}{t}$ normalized remaining storage, then $e \geq t$.} $e$ is increased by at least $1$ after this iteration. This completes the proof of Lemma \ref{lm: 3}.
\end{IEEEproof}

By Lemmas \ref{lm: 2} and \ref{lm: 3}, we can conclude that at most $t-1$ PFs and $N-t$ CFs are possible during the execution of Algorithm \ref{algorithm:1} as $N'$ is decreased from $N$ to $t$. Then when $N'=t$, there are $t$ DBs with equal remaining storage and the special case of Algorithm \ref{algorithm:1} fills the remaining storage of these DBs. As a result, at most $(t-1) + (N-t) + 1 = N$ iterations of Algorithm \ref{algorithm:1} are necessary to completely fill the available storage at the DBs.
\end{IEEEproof}
\begin{remark}
  It can also be shown that if $N'\geq 2t$ then an iteration will result in a CF. In other words, the first $N-2t+1$ iterations are guaranteed to be a CF.
  \end{remark}
  \begin{remark} 
  If $m[N-N'+1] = \frac{t'}{t}-m[N-t+1]$, then the iteration will result in a CF and $e$ will increase by at least $1$. Moreover, if there are multiple nodes with normalized remaining storage equal to $m[N-t+1]$, then $e$ will increase by more than $1$ if the iteration is a PF. These special cases demonstrate that in some cases a number of iterations strictly less than $N$ may be sufficient to fills the DBs.
\end{remark}

\section{A Capacity Achieving Placement Design for Non-Integer $t$}
\label{sec: nonintt}
In previous sections, we assumed that $t$ is an integer and sub-message sets are always stored at $t$ nodes. In practice, $t$ may not be an integer. In this section, we aim to find a capacity achieving solution to the general heterogeneous SC-PIR problem with a non-integer $t$. The main challenge is to design a {\em storage sharing} scheme that satisfies (\ref{eq: condition FP}) which guarantees the existence of a FP solution.
The key idea is to split the storage placement problem into two storage placement sub-problems with integer $t$. In the following, we first provide a motivating example and then derive the sufficient conditions to meet SC-PIR capacity. A general scheme that meets the sufficient conditions is presented at the end of this section.

\subsection{An Example for $t\notin \mathbb{Z}^+$}

Let $N=4$ with storage requirements defined by
\be
\boldsymbol{\mu} = \left[\frac{1}{5}, \;\; \frac{1}{5}, \;\; \frac{2}{5}, \;\; \frac{3}{5}, \;\; 1\right]
\ee
and $t = \frac{12}{5}$. Algorithm \ref{algorithm:1} cannot operate on these requirements since $t$ is not an integer. Instead we have to satisfy the conditions (\ref{eq: notintt1}) and (\ref{eq: notintt2}) of Lemma \ref{th: suff_cond}. In other words, for each DB we split the storage into two parts. We define $\boldsymbol{\mu}^{(2)}$ and $\boldsymbol{\mu}^{(3)}$ as the set of storage requirements allotted for the $\left(\boldsymbol{\mu}^{(2)},2\right)$ and $\left(\boldsymbol{\mu}^{(3)},3\right)$ FPs, respectively, such that $\boldsymbol{\mu}^{(2)} + \boldsymbol{\mu}^{(3)} = \boldsymbol{\mu}$.

From (\ref{eq: notintt1}) and (\ref{eq: notintt2}) of Lemma \ref{th: suff_cond}, the elements of $\boldsymbol{\mu}^{(2)}$ must sum to $\lfloor  t \rfloor(\lceil t \rceil - t) = \frac{6}{5}$ and the elements of $\boldsymbol{\mu}^{(3)}$ must sum to $\lceil t \rceil(t-\lfloor t \rfloor)=\frac{6}{5}$. A naive approach is to simply split the storage of each DB in half:
\be
\boldsymbol{\mu}^{(2)} = \boldsymbol{\mu}^{(3)} = \frac{1}{2}\boldsymbol{\mu} = \left[\frac{1}{10}, \;\; \frac{1}{10}, \;\; \frac{1}{5}, \;\; \frac{3}{10}, \;\; \frac{1}{2}\right].
\ee
While (\ref{eq: notintt1}) and (\ref{eq: notintt2}) are met, there is an issue with this approach since a $\left(\boldsymbol{\mu}^{(3)},3\right)$-FP solution does not exist. This is the case since $\mu^{(3)}[4] = \frac{1}{2} > \frac{1}{3}\sum_{n=1}^{4}\mu^{(3)} [n]= \frac{2}{5}$ and condition (\ref{eq: condition FP}) of Theorem \ref{th: exist} is not met.

We can use an alternative approach to define the storage sharing in order to meet Theorem \ref{th: exist} by
enforcing $\mu^{(2)}[4] \geq \frac{3}{5}$. Then we find
\begin{align}
&\mu^{(3)}[4] + \mu^{(2)}[4]= \mu [4] = 1, \\
&\mu^{(3)}[4]\leq 1- \frac{3}{5} = \frac{2}{5}.
\end{align}
To ensure that there are no issues with the storage sharing of the other DBs, we enforce
\begin{align}\label{eq: nonintt_b1}
    \mu^{(2)}[n]\geq \mu [n] -  \frac{\lceil  t  \rceil(t-\lfloor t \rfloor)}{\lceil  t  \rceil} = \mu [n] - \frac{2}{5}, \text{ for } n=1,2,3,4
\end{align}
such that $\mu^{(3)}[n] = \mu[n] - \mu^{(2)}[n] \leq \frac{2}{5}$ which ensures that $\boldsymbol{\mu}^{(3)}$ has a valid FP solution. Similarly, we enforce
\be \label{eq: nonintt_b2}
\mu^{(3)}[n]\geq \mu [n] - \frac{\lfloor t \rfloor(\lceil  t  \rceil - t )}{\lfloor t \rfloor} = \mu [n] - \frac{3}{5}, \text{ for } n=1,2,3,4
\ee
so that $\boldsymbol{\mu}^{(2)}$ has a valid FP solution. 
Using this approach we can define the data sharing scheme by
\begin{align}
  \boldsymbol{\mu}^{(2)} & = \left[ 0, \;\; 0, \;\; 0, \;\; \frac{1}{5}, \;\; \frac{3}{5}  \right] +  \left[ \frac{1}{15}, \;\; \frac{1}{15}, \;\; \frac{2}{15}, \;\; \frac{2}{15}, \;\; 0 \right]\label{eq: mu_1} \\
  & = \left[ \frac{1}{15}, \;\; \frac{1}{15}, \;\; \frac{2}{15}, \;\; \frac{1}{3}, \;\; \frac{3}{5}  \right]\\
  \boldsymbol{\mu}^{(3)} & = \left[ 0, \;\; 0, \;\; 0, \;\; 0, \;\; \frac{2}{5} \right] + \left[ \frac{2}{15}, \;\; \frac{2}{15}, \;\; \frac{4}{15}, \;\; \frac{4}{15}, \;\; 0  \right] \label{eq: mu_2} \\
  & = \left[ \frac{2}{15}, \;\; \frac{2}{15}, \;\; \frac{4}{15}, \;\; \frac{4}{15},  \;\; \frac{2}{5} \right].
\end{align}
Specifically, on the RHS of (\ref{eq: mu_1}) and (\ref{eq: mu_2}), the first term ensures Theorem \ref{th: exist} is met. In our general scheme, we label these terms as $\boldsymbol{m}_1$ and $\boldsymbol{m}_2$, respectively. The second term on the RHS of (\ref{eq: mu_1}) and (\ref{eq: mu_2}) is defined such that both of the resulting vectors sum to $\frac{6}{5}$ and (\ref{eq: notintt1}) and (\ref{eq: notintt2}) are met which are also defined on our general scheme. At this point, Algorithm \ref{algorithm:1} can be used to define the placement of the equivalent $\left(\boldsymbol{\mu}^{(2)},2\right)$ and $\left(\boldsymbol{\mu}^{(3)},3\right)$ FPs.
\subsection{Storage Sharing Sufficient Conditions}

Define $\boldsymbol{\mu}^{(\lfloor t \rfloor)}, \boldsymbol{\mu}^{(\lceil t \rceil)}\in\mathbb{R}^N_+$ such that $\boldsymbol{\mu}^{(\lfloor t \rfloor)} + \boldsymbol{\mu}^{(\lceil t \rceil)} = \boldsymbol{\mu}$,
\be \label{eq: cond1}
\sum_{i=1}^{N}\mu^{(\lfloor t \rfloor)}[i] = \lfloor t \rfloor(\lceil t \rceil - t),
\ee
\be\label{eq: cond2}
\sum_{i=1}^{N}\mu^{(\lceil t \rceil)}[i] = \lceil t \rceil(t - \lfloor t \rfloor),
\ee
and the condition (\ref{eq: condition FP}) for $\mu^{(\lfloor t \rfloor)}[n] $ and $\mu^{(\lceil t \rceil)}[n] $ is given by
\be\label{eq: cond3}
\mu^{(\lfloor t \rfloor)}[n] \leq \frac{\sum_{i=1}^{N}\mu^{(\lfloor t \rfloor)}[i]}{\lfloor t \rfloor} = \lceil t \rceil - t,
\ee
for all $n \in [N]$ and
\be\label{eq: cond4}
\mu^{(\lceil t \rceil)}[n] \leq \frac{\sum_{i=1}^{N}\mu^{(\lceil t \rceil)}[i]}{\lceil t \rceil} = t - \lfloor t \rfloor,
\ee
for all $n \in [N]$. We find that (\ref{eq: cond1}) and (\ref{eq: cond2}) satisfy (\ref{eq: notintt1}) and (\ref{eq: notintt2}) to achieve heterogeneous SC-PIR capacity. Moreover, (\ref{eq: cond3}) and (\ref{eq: cond4}) guarantee that a solution exists to both the ($\boldsymbol{\mu}^{(\lfloor t \rfloor)},\lfloor t \rfloor$)-FP and ($\boldsymbol{\mu}^{(\lceil t \rceil)},\lceil t \rceil$)-FP. Then, split each message $W_k$ into two disjoint sub-messages, $W_k^{(\lfloor t \rfloor)}$ of size $(\lceil t \rceil - t)L$ bits and $W_k^{(\lceil t \rceil)}$ of size $(\lfloor t \rfloor - t)L$ bits which are used to for each FP. These two FPs can then be solved by Algorithm \ref{algorithm:1}.

\subsection{A Storage Sharing Solution}

Given $\boldsymbol{\mu}$, the following process will yield a valid $\boldsymbol{\mu}^{(\lfloor t \rfloor)}$ and $ \boldsymbol{\mu}^{(\lceil t \rceil)}$ which meet the above conditions. Define $\boldsymbol{m}_1,\boldsymbol{m}_2\in\mathbb{R}_+^N$ such that
\be \label{eq: m1}
m_1[n] = \Big[ \mu[n] - ( t - \lfloor t \rfloor) \Big]^+,
\ee
for all $n\in[N]$ and
\be \label{eq: m2}
m_2[n] = \Big[ \mu[n] - (\lceil t \rceil - t) \Big]^+,
\ee for all $n \in [N]$, where $[\cdot]^+$ returns the input if the input is non-negative, or returns $0$ otherwise. Let
\be \label{eq: r}
r = \frac{\lfloor t \rfloor(\lceil t \rceil - t) - \sum_{n=1}^{N}m_1[n]}{t - \sum_{n=1}^{N}m_1[n] - \sum_{n=1}^{N} m_2[n]},
\ee
then let
\be \label{eq: mu1}
\boldsymbol{\mu}^{(\lfloor t \rfloor)} = \boldsymbol{m}_1 + (\boldsymbol{\mu} - \boldsymbol{m}_1 - \boldsymbol{m}_2)\cdot r
\ee
and
\be\label{eq: mu2}
\boldsymbol{\mu}^{(\lceil t \rceil)} = \boldsymbol{m}_2 + (\boldsymbol{\mu} - \boldsymbol{m}_1 - \boldsymbol{m}_2)\cdot (1-r).
\ee

The correctness of this scheme for $t\notin \mathbb{Z}^+$ is proved in Appendix \ref{sec: corr_app}. Note that, the memory allocation of $\boldsymbol{m}_1$ and $\boldsymbol{m}_2$ are required to have a valid solution, there are many design choices to define the remaining memory sharing. Here, we provide one approach which is to split the remaining memory of each DB with constant ratio $r$.

\section{Discussion}
\label{sec: Discussion}
Recent works on SC-PIR suggest that coded caching \textit{meets} PIR \cite{attia2018capacity,tandon2018pir}; that is, the file placement solutions of coded caching \cite{maddah2014fundamental} are useful for the SC-PIR sub-message placement problem. In this work, we show that coded caching placement techniques are not necessary for SC-PIR by proposing two novel sub-message placement schemes which achieve the capacity. In the coded caching problem, assigning different files to an exponentially large number of overlapping user groups is necessary to create multicasting opportunities such that a user can cancel ``interference" from a received coded transmission which also serves other users. The SC-PIR problem is less complex in that only one user is being served. In fact, as was demonstrated with our first proposed homogeneous scheme, it is not necessary for the sub-message placement groups to overlap at all. Moreover, the file (or sub-message) placement paradigms of coded caching and SC-PIR are inherently different. In coded caching, files are being placed among users that wish to download content, while in SC-PIR, sub-messages are being placed among databases which are serving one user's request. Therefore, it is not surprising the two problems could have different solutions for the storage/file placement problem.

The results of Section \ref{sec: conv} demonstrated that Algorithm \ref{algorithm:1} requires at most $N$ iterations to complete. Since each iteration defines one sub-message per message, the number of sub-messages per message resulting from Algorithm \ref{algorithm:1} is at most $N$. This leads to the following corollary.
\begin{corollary}
Given a set of storage requirements $\boldsymbol{\mu} \in \mathbb{R}^N_+$ such that $\mu[n]\leq 1$ for all $n\in[N]$, $t\in\mathbb{Z}^+$ and $t\geq 1$, there exists a capacity achieving heterogeneous SC-PIR scheme with at most $N X_{d}$ sub-messages per message where $X_d$ is the required number of sub-messages for the FS-PIR delivery scheme.
\hfill$\square$
\end{corollary}

Surprisingly, from homogeneous to heterogeneous SC-PIR, there is no loss in rate as shown in \cite{banawan2019capacity} and no increase in the number of sub-messages as shown here.\footnote{Notice that we have mainly discussed the number of sub-messages which result from the storage placement and not the number of sub-message for the delivery phase.} The total number of sub-messages is the product of the number of sub-messages necessary for the storage and delivery phases. By using the recent result of \cite{tian2018capacity} for delivery, the total number of sub-messages per message is $N\times(N-1) < N^2$. Amazingly, this implies that heterogeneous SC-PIR may be practical for a large number of DBs. Furthermore, the number of sub-messages is constant with respect to the number of messages, $K$.


Another important aspect is the required message size in terms of the number of bits using Algorithm~\ref{algorithm:1} for the general heterogeneous SC-PIR problem. 
In this case, the sub-messages have different sizes 
and $\alpha[f]L$ must be an integer for all $f\in[F]$. 
In general, the minimum size of $L$ based on Algorithm~\ref{algorithm:1} is still $O(N^2)$. 
However, it appears to be a function of all the distinct values of $\boldsymbol{\mu}$.

It is also possible to use Algorithm \ref{algorithm:1} on a set of homogeneous storage requirements. It is interesting to observe how the result compares to the explicit homogeneous schemes presented here. It can be shown that Algorithm \ref{algorithm:1} will completely fill $t$ DBs with each iteration until the number of remaining DBs is $N' \leq 2t-1$. This pattern reflects the storage placement of SC-PIR scheme for $\frac{N}{t}\in \mathbb{Z}^+$. In fact if $\frac{N}{t}\in \mathbb{Z}^+$, then Algorithm $1$ will yield the same storage placement as our first homogeneous SC-PIR scheme. Otherwise, once $N' \leq 2t-1$, Algorithm \ref{algorithm:1} will result in a cyclic placement mimicking our second proposed homogeneous SC-PIR scheme.

This work presents several interesting directions for future work. First, it remains an open problem to determine the minimum  message size $L$ for a given set of SC-PIR parameters. It was shown in \cite{tian2018capacity} that the minimum $L$ of an FS-PIR problem can be reduced significantly from $N^{K-1}$ in {\cite{sun2017optimal}} to $N-1$. The new FS-PIR scheme \cite{tian2018capacity} and proof techniques therein may be useful to derive the minimum $L$ for the homogeneous and heterogeneous SC-PIR problems.
Second, another work \cite{wei2018capacity} has considered random placement among  databases  where a database stores a bit of a given message with probability $\mu$. Interestingly, this placement method was also used in \cite{maddah2015decentralized} for the coded caching problem. It will be meaningful to examine alternative random placement strategies for the SC-PIR problem where messages are split into a finite number of sub-messages.

\section{Conclusion}
\label{sec: Conclusion}
In this work, we proposed novel designs of both homogeneous and heterogeneous SC-PIR such that the capacity can be achieved. The SC-PIR schemes were developed from scratch and surprisingly we find that our schemes require only a polynomial number of sub-messages per message. Moreover, we provided necessary and sufficient conditions to achieve capacity for SC-PIR which can aid the design of further SC-PIR schemes. These results not only proved that the general storage problem for heterogeneous SC-PIR has a solution, but also the existence of a simple iterative storage placement algorithm such that the conditions are met after each iteration. In addition, when $t$ is an integer, we also showed that the proposed iterative algorithm converges within $N$ iterations. Finally, the algorithm was extended to account for non-integer $t$. 

\appendices
\section{Privacy of General Design Architecture}
\label{sec: priv_pf}

We aim to prove the condition of (\ref{eq: PIR 3}) holds when using the design architecture of Section \ref{sec: des_arch}. Let
\begin{align}
Q_n^{[k]} &= Q_{n,1}^{[k]}, \ldots , Q_{n,F}^{[k]} \\
A_n^{[k]} &= A_{n,1}^{[k]}, \ldots , A_{n,F}^{[k]}
\end{align}
where $Q_{n,f}^{[k]}$ and $A_{n,f}^{[k]}$ are the query to and response from DB $n$ when the user is privately downloading the sub-message $W_{k,f}$ of the requested message, $W_k$.\footnote{Note that, it is possible that the realization of $Q_{n,f}^{[k]}$ and $A_{n,f}^{[k]}$ are empty because DB $n$ does not have $W_{k,f}$ locally stored.} Furthermore, let
\begin{align}
Z_n = Z_{n,1} , \ldots ,  Z_{n,F}
\end{align}
where $Z_{n,f}$ is the storage contents of DB $n$ which intersects the sub-message set $M_f = W_{1,f},\ldots , W_{K,f}$. Finally, let
\be
V_{n,f} = Q_{n,f}^{[k]}, A_{n,f}^{[k]}, W_{1,f}, \ldots, W_{K,f}, Z_{1,f}, \ldots, Z_{N,f}.
\ee
The LHS of (\ref{eq: PIR 3}) can be re-written as
\begin{align}
    I&\left( k ;  Q_n^{[k]}, A_n^{[k]}, W_1, \ldots, W_K, Z_1, \ldots, Z_N \right) \\
    &= I\left( k ; V_{n,1},\ldots , V_{n,F} \right) \\
    &= \sum_{f=1}^{F} I\left( k ; V_{n,f} | V_{n,1}, \ldots , V_{n,f-1} \right). \label{eq: priv_pf_sum}
\end{align}
By showing each term of (\ref{eq: priv_pf_sum}) is $0$, we can prove the condition of (\ref{eq: PIR 3}) holds. The first term
\begin{align}
I&\left( k ; V_{n,1} \right) \\
&= I\left( k ; Q_{n,1}^{[k]}, A_{n,1}^{[k]}, W_{1,1}, \ldots, W_{K,1}, Z_{1,1}, \ldots, Z_{N,1} \right)
\end{align}
equals $0$ because we assume the FS-PIR scheme used in private. Next, we find
\begin{align}
I&\left( k ; V_{n,f} | V_{n,1}, \ldots , V_{n,f-1} \right) \\
&\stackrel{\text{(a)}}{=} H\left( V_{n,f} | V_{n,1}, \ldots , V_{n,f-1} \right) \nonumber\\
&\;\;\;\;\;\;- H\left( V_{n,f} | k, V_{n,1}, \ldots , V_{n,f-1} \right)\\
&\stackrel{\text{(b)}}{=} H\left( V_{n,f} | V_{n,1}, \ldots , V_{n,f-1} \right) \nonumber\\
&\;\;\;\;\;\;-H\left(   V_{n,1}, \ldots , V_{n,f} | k \right) + H\left(  V_{n,1}, \ldots , V_{n,f-1} | k \right) \\
&\stackrel{\text{(c)}}{=} H\left( V_{n,f} | V_{n,1}, \ldots , V_{n,f-1} \right) \nonumber\\
&\;\;\;\;\;\;-H\left(   V_{n,1}, \ldots , V_{n,f}  \right) + H\left(  V_{n,1}, \ldots , V_{n,f-1}  \right) \\
&\stackrel{\text{(d)}}{=} 0
\end{align}
where (a), (b) and (d) hold from rules of information theory. Moreover, (c) holds because we assume we are using a private FS-PIR scheme for query generation such that the distribution of the query, answers, messages and storage contents are independent of the label of the desired message, $k$.

\section{Proof of FP Solution Existence}
\label{sec: FPexistpf}
\begin{IEEEproof}
The proof is split into two claims.
\begin{claim}\label{cl: exist_pf1}
  If a $(\boldsymbol{m},\tau)$-FP exists then $m[n] \leq \frac{\sum_{i=1}^{N}m[i]}{\tau}$ for all $n \in [N]$.
\end{claim}

The proof of Claim \ref{cl: exist_pf1} is as follows. Define a set $\mathcal{B} \subset \mathbb{R}_+^N$ such that $\mathcal{B}$ includes all possible $\{0,1\}$-vectors with exactly $\tau$ $1$s. A $(\boldsymbol{m},\tau)$-FP solution exists if and only if $\boldsymbol{m} = \sum_{\boldsymbol{b} \in \mathcal{B}} \alpha_{\boldsymbol{b}} \boldsymbol{b}$
where $\alpha_{\boldsymbol{b}} \in \mathbb{R}_+$ for all $\boldsymbol{b} \in \mathcal{B}$. We perform the following inductive process on the basis $\mathcal{B}$. First, define $\boldsymbol{m}^{(0)} = \boldsymbol{0}\in\mathbb{R}_+^N$. It is clear that
\be
m^{(0)}[n] \leq \frac{\sum_{i=1}^{N}m^{(0)}[i]}{\tau}
\ee
for all $n \in [N]$.
Next, define some order to the vectors of $\mathcal{B}$ such that $\mathcal{B} = \left\{\boldsymbol{b}^{(1)}, \boldsymbol{b}^{(2)}, \ldots ,  \boldsymbol{b}^{(|\mathcal{B}|)} \right\}$. Given some $\boldsymbol{m}^{(k)} \in \mathbb{R}_+^N$ such that $m^{(k)}[n] \leq \frac{\sum_{i=1}^{N}m^{(k)}[i]}{\tau}$ for all $n \in [N]$, let
\be
\boldsymbol{m}^{(k+1)} = \boldsymbol{m}^{(k)} + \alpha_{\boldsymbol{b}^{(k+1)}} \boldsymbol{b}^{(k+1)}
\ee
then,
\begin{align}
\max_n m^{(k+1)}[n] &\leq \alpha_{\boldsymbol{b}^{(k+1)}} + \max_i m^{(k)}[n] \\
&\leq \alpha_{\boldsymbol{b}^{(k+1)}} + \frac{\sum_{i=1}^{N}m^{(k)}[i]}{\tau} \\
&= \frac{\tau \alpha_{\boldsymbol{b}^{(k+1)}} + \sum_{i=1}^{N}m^{(k)}[i]}{\tau} \\
&= \frac{\sum_{i=1}^{N}\alpha_{\boldsymbol{b}^{(k+1)}} b^{(k+1)}[i] + m^{(k)}[i]}{\tau} \\
&= \frac{\sum_{i=1}^{N}m^{(k+1)}[i]}{\tau}.
\end{align}
When $k+1 = |\mathcal{B}|$, then $\boldsymbol{m} = \boldsymbol{m}^{(k+1)}$ and therefore $m[n] \leq \frac{\sum_{i=1}^{N}m[i]}{\tau}$ for all $n \in [N]$. This completes the proof of Claim \ref{cl: exist_pf1}.

To complete the proof of Theorem \ref{th: exist}, we prove the following claim.
\begin{claim}\label{cl: exist_pf2}
  If $m[n] \leq \frac{\sum_{i=1}^{N}m[i]}{\tau}$ for all $n \in [N]$ then a $(\boldsymbol{m},\tau)$-FP solution exists.
\end{claim}

The proof of Claim \ref{cl: exist_pf2} is as follows. Given some $a \in \mathbb{R}_+$, the set
\begin{align}
\mathcal{M}_a = \Big \{ \boldsymbol{m}' \in \mathbb{R}_+^N : & \sum_{i=1}^{N}m'[i]= a,\nonumber\\
 &m'[n] \leq \frac{\sum_{i=1}^{N}m'[i]}{\tau} \text{ for all } n \in [N] \Big \}
\end{align}
is defined by the intersection of $2N$ half-spaces and $1$ plane and therefore $\mathcal{M}_a$ is convex. Moreover, $\mathcal{M}_a$ is bounded and closed because $0\leq m'[n] \leq \frac{a}{\tau}$ for all $n \in [N]$. Therefore, $\mathcal{M}_a$ can be defined by the set of all convex combinations of the corner points of $\mathcal{M}_a$, labeled as $  \mathcal{C}_a$. In other words,
\be
\mathcal{M}_a = \left \{ \sum_{\boldsymbol{c} \in \mathcal{C}_a}\lambda[i] \boldsymbol{c} \ : \boldsymbol{\lambda} \in \Delta_{|\mathcal{C}_a|}   \right \}
\ee
where $\Delta_{|\mathcal{C}_a|}$ is the unit simplex of dimension $|\mathcal{C}_a|$.

The corner points, $\mathcal{C}_a$, are defined by the intersections of the planes that define the set $\mathcal{M}_a$. Given an integer $\tau'$ such that $0\leq \tau ' \leq N$, and some set $\mathcal{S} \subseteq [N]$ such that $|\mathcal{S}|=\tau '$. Now, consider the set of planes defined by $m'[n] = \frac{\sum_{i=1}^{N}m'[i]}{\tau} = \frac{a}{\tau}$ for all $n\in \mathcal{S}$. Then,
\begin{align}
\sum_{n \in [N]}m'[n] &=\sum_{n \in \mathcal{S}}m'[n] +  \sum_{n \in [N] \setminus\mathcal{S}}m'[n] \geq \sum_{n \in \mathcal{S}}m'[n] = \tau'\cdot\frac{a}{\tau}.
\end{align}
If $\tau ' > \tau$ then $\sum_{n \in [N]}m'[n] > a$ and the intersection of the $\tau '$ planes is not included in $\mathcal{M}_a$. If $\tau ' = \tau$, then $\sum_{n \in [N]}m'[n] \geq a$ and equality holds if and only if $\sum_{n \notin \mathcal{S}}m'[n] = 0$, and furthermore, since $m'[n] \geq 0$ for all $n\in [N]$, this yields a corner point $m'[n] = \frac{a}{\tau}$ if $n \in\mathcal{S}$, and $m'[n] = 0$ if $n \in[N]\setminus\mathcal{S}$. Finally, if $\tau ' < \tau$, then $\sum_{n \in \mathcal{S}}m'[n]  < a$. To define a point in $\mathcal{M}_a$, some $m[n]$ for $n\in[n]\setminus \mathcal{S}$ must be non-zero and to find a corner point we intersection planes of the form $m[n] = \frac{a}{\tau}$. However, eventually, we find that we are ultimately intersecting $\tau$ planes of the form $m[n]=\frac{a}{\tau}$. These corner points were already included when $\tau ' = \tau$. Hence,
\begin{align}
\mathcal{C}_a = \Big \{ \boldsymbol{m}' \in \mathbb{R}_+^N : m'[n] = \frac{a}{\tau} \text{ if } n \in \mathcal{S}, m'[n] = 0 \text{ if } n \in [N]\setminus\mathcal{S}, \mathcal{S}\subseteq [N] , |\mathcal{S}| = \tau  \Big\}.
\end{align}

In fact,
\be
\mathcal{C}_a = \left\{ \frac{a}{\tau}\boldsymbol{b} : \boldsymbol{b} \in \mathcal{B}  \right\}
\ee
where $\mathcal{B}$ is the basis defined in the proof of Claim \ref{cl: exist_pf1}. Therefore,
\be
\mathcal{M}_a = \left \{ \frac{a}{\tau}\sum_{\boldsymbol{b} \in \mathcal{B}}\lambda[i] \boldsymbol{b} \ : \boldsymbol{\lambda} \in \Delta_{|\mathcal{B}|} \right \}
\ee
and for every point $\boldsymbol{m}'\in\mathcal{M}_a$, there exists a $(\boldsymbol{m}',\tau)$-FP solution. This holds for all $a \geq 0$. This completes the proof of Claim \ref{cl: exist_pf2}.
\end{IEEEproof}

\section{Correctness of Algorithm \ref{algorithm:1}}
\label{sec: correct_alg}
In the following, we demonstrate that each iteration fills a non-zero, positive amount of storage. WLOG we assume $m[1] \leq m[2] \leq \ldots \leq m[N]$. Furthermore, assuming that $m[N-N'+1] > 0$ and a ($\boldsymbol{m},t$)-FP solution exists such that $m[N-t+1] \leq \frac{\sum_{i=1}^{N}m[i]}{t} = \frac{t'}{t}$, then observing (\ref{eq: alpha}), we can see that $\alpha \geq 0$. Moreover, $\alpha=0$ if and only if
\be
m[N-t+1] = \frac{t'}{t} = \frac{\sum_{i=1}^{N}m[i]}{t}
\ee
and in this case we find for all $n\in [N-t+1:N]$ that
\be
\frac{\sum_{i=1}^{N}m[i]}{t} = m[N-t+1] \leq  m[n]  \leq \frac{\sum_{i=1}^{N}m[i]}{t}.
\ee
and $m[n] = m[N-t+1]$. This means that $N'=t$ and each of the $t$ DBs has the same amount of remaining storage. In this case, $\alpha = m[N-t+1]$ as defined by the exception when $N'=t$.

Next, we demonstrate that after an iteration the remaining storage among the DBs is such that a FP solution exists. Let
\be
\boldsymbol{m}' = \boldsymbol{m} - \alpha\cdot [\underbrace{ 0 , \ldots , 0}_{N-N'},1, \underbrace{ 0 , \ldots , 0}_{N'-t} ,\underbrace{ 1 , \ldots , 1}_{t-1} ]
\ee
represent the remaining storage after a particular iteration. Note that, the elements of $\boldsymbol{m}'$ are not necessarily in order. After an iteration, the largest remaining storage at any node is either $m'[N]~=~m[N] - \alpha$ or $ m'[N-t+1]= m[N-t+1]$. Assuming a ($\boldsymbol{m},t$)-FP solution exist, then
 \begin{align}
 m'[N] = m[N] - \alpha \leq \frac{\sum_{i=1}^{N}m[i]}{t} - \alpha =  \frac{\sum_{i=1}^{N}m'[i]}{t}.
 \end{align}
Also, by (\ref{eq: alpha}), $\alpha \leq \frac{\sum_{i=1}^{N}m[i]}{t} - m[N-t+1]$ and $m'[N~-~t~+~1]~=~m[N-t+1]$, then
\be
m'[N-t+1] \leq \frac{\sum_{i=1}^{N}m[i]}{t} - \alpha =  \frac{\sum_{i=1}^{N}m'[i]}{t}.
\ee
Furthermore, $\alpha \leq m[N-N'+1] \leq m[n]$ and $m'[n] \geq m[n] - \alpha \geq 0$ for all $n\in [N-N'+1:N]$. Finally $m'[n] = m[n]=0$ for all $n\in[1:N-N']$. Since $0 \leq m'[n] \leq \frac{\sum_{i=1}^{N}m'[i]}{t}$ for all $n\in [N]$, by using Theorem~\ref{th: exist}, a ($\boldsymbol{m}',t$)-FP solution exists.

\section{Correctness of Non-Integer $t$ Scheme}
\label{sec: corr_app}
In this section, when $t$ is not a integer, we will show that $\boldsymbol{\mu}^{(\lfloor t \rfloor)}$ and $ \boldsymbol{\mu}^{(\lceil t \rceil)}$ as defined by (\ref{eq: mu1}) and (\ref{eq: mu2}), respectively, are non-negative vectors which satisfy the conditions of (\ref{eq: cond1})-
(\ref{eq: cond4}). In the following, we show (\ref{eq: cond1}) is satisfied.
\begin{align}
\sum_{n=1}^{N}\mu^{(\lfloor t \rfloor)}[n] &= \sum_{n=1}^{N}m_1[n] + r\left( t - \sum_{n=1}^{N}m_1[n] - \sum_{n=1}^{N} m_2[n] \right) = \sum_{n=1}^{N}m_1[n] + \lfloor t \rfloor(\lceil t \rceil - t) - \sum_{n=1}^{N}m_1[n] \nonumber \\
&= \lfloor t \rfloor(\lceil t \rceil - t).
\end{align}
In the following, we show (\ref{eq: cond2}) is satisfied.
\begin{align}
\sum_{n=1}^{N}\mu^{(\lfloor t \rfloor)}[n] &= \sum_{n=1}^{N}m_2[n] + (1-r)\left( t - \sum_{n=1}^{N}m_1[n] - \sum_{n=1}^{N} m_2[n] \right) \nonumber\\
&=\sum_{n=1}^{N}m_2[n] + t - \sum_{n=1}^{N}m_1[n] - \sum_{n=1}^{N} m_2[n] - \lfloor t \rfloor(\lceil t \rceil - t) + \sum_{n=1}^{N}m_1[n] \nonumber\\
&= t - \lfloor t \rfloor(\lceil t \rceil - t)
= \lceil t \rceil(t-\lfloor t \rfloor).
\end{align}

Next, we use the following lemmas which are proven in the latter part of Appendix \ref{sec: corr_app}.

\begin{lemma} \label{cl: 3}
  Given the vectors $\boldsymbol{m}_1$ and $\boldsymbol{m}_2$ defined in (\ref{eq: m1}) and (\ref{eq: m2}), respectively, we have
  \be
  \label{eq: lemma 3}
   m_1[n] + m_2[n] \leq \mu [n]
   \ee for all $n\in[N]$. Moreover, equality holds, $m_1[n] + m_2[n] = \mu[n]$, {\it if and only if} $\mu[n] \in \{0, 1\}$.\footnote{Note that, when $\mu[n] \in \{0, 1\}$ for all $n\in[N]$, $t$ is an integer, which is not the scenario of interest in this section.}
\hfill$\square$
\end{lemma}

\begin{lemma} \label{cl: 4}
  Given $r$ as defined in (\ref{eq: r}), we have $0\leq r < 1$.
 \hfill $\square$
\end{lemma}

Given Lemmas \ref{cl: 3} and \ref{cl: 4}, since $ m_1[n]\geq 0$ and $ m_2[n]\geq 0$ for all $n\in [N]$, then $\boldsymbol{\mu}^{(\lfloor t \rfloor)}$ and $ \boldsymbol{\mu}^{(\lceil t \rceil)}$ have only non-negative values. Moreover,
\begin{align}
\mu^{(\lfloor t \rfloor)}[n] &< m_1[n] + (\mu[n] - m_1[n] - m_2[n]) = \mu[n]  - m_2[n] = \mu[n] - \Big[ \mu[n] - (\lceil t \rceil - t) \Big]^+ \leq \lceil t \rceil - t
\end{align}
for all $n\in[N]$. Hence, (\ref{eq: cond3}) is satisfied. Similarly,
\begin{align}
  \mu^{(\lceil t \rceil)}[n] &\leq m_2[n] + (\mu[n] - m_1[n] - m_2[n]) = \mu[n] - m_1[n] = \mu[n] - \Big[ \mu[n] - (t-\lfloor t \rfloor) \Big]^+ \leq t-\lfloor t \rfloor
\end{align}
for all $n\in[N]$  such that (\ref{eq: cond4}) is satisfied. This completes the proof of correctness. The rest of this Appendix \ref{sec: corr_app} is devoted to proving Lemmas \ref{cl: 3} and \ref{cl: 4}.

\subsection{Proof of Lemma \ref{cl: 3}}

We first prove (\ref{eq: lemma 3}). In the following, according to the value of $\mu[n]$, we have four cases.
\begin{itemize}
\item If $\mu [n] \leq t - \lfloor t \rfloor$ and $\mu [n] \leq \lceil t \rceil - t$, then
\be
m_1[n] = m_2[n] = 0
\ee
 and
  \be \label{eq: cl3_1}
  m_1[n] + m_2[n] = 0 \leq \mu [n].
   \ee
\item If $\mu [n] > t - \lfloor t \rfloor$ and $\mu [n] \leq \lceil t \rceil - t$, then
\be
m_1[n] = \mu[n] - (t - \lfloor t \rfloor),
\ee
\be
m_2[n] = 0
\ee
and \be \label{eq: cl3_2}
m_1[n] + m_2[n] = \mu[n] - (t - \lfloor t \rfloor) < \mu [n].
\ee

\item If $\mu [n] \leq t - \lfloor t \rfloor$ and $\mu [n] > \lceil t \rceil - t$, then
\be \label{eq: cl3_3}
m_1[n] + m_2[n] = \mu[n] - (\lceil t \rceil - t) < \mu [n].
\ee
\item If $\mu [n] > t - \lfloor t \rfloor$ and $\mu [n] > \lceil t \rceil - t$, then
\be \label{eq: cl3_4}
m_1[n] + m_2[n] \buildrel{(a)} \over = 2\mu[n] - 1 \buildrel{(b)} \over\leq \mu[n],
\ee
where (a) is because $(t - \lfloor t \rfloor) + (\lceil t \rceil - t) = 1$ and (b) is because $\mu[n]\leq 1$.
\end{itemize}

We prove the last part of Lemma \ref{cl: 3} as follows. By observing (\ref{eq: cl3_1}), (\ref{eq: cl3_2}), (\ref{eq: cl3_3}) and (\ref{eq: cl3_4}), $m_1[n] + m_2[n] = \mu[n]$ if $\mu[n]=0$, as shown in (\ref{eq: cl3_1}), or if $\mu[n]=1$ as shown in (\ref{eq: cl3_4}), and otherwise $m_1[n] + m_2[n] \neq \mu[n]$. Therefore, $m_1[n] + m_2[n] = \mu[n]$ {\it if and only if} $\mu[n] \in \{0, 1\}$. This completes the proof of Lemma \ref{cl: 3}.

\subsection{Proof of Lemma \ref{cl: 4}}

First, 
we show that the denominator of (\ref{eq: r}) is strictly positive. By Lemma \ref{cl: 3}, $ m_1[n] + m_2[n] \leq \mu [n]$ for all $n\in [N]$, therefore
\begin{align}
t - \sum_{n=1}^{N}&m_1[n] - \sum_{n=1}^{N} m_2[n] = \sum_{n=1}^{N} (\mu[n] - m_1[n] - m_2[n]) \geq 0. \label{eq: r_dem}
\end{align}
Furthermore, equality holds in (\ref{eq: r_dem}) {\it if and only if} $\mu[n] = m_1[n] + m_2[n]$ for all $n\in[N]$. By Lemma \ref{cl: 3}, we obtain $\mu[n] \in\{ 0,1 \}$ for all $n\ \in [N]$, which means that in this case $t$ is an integer (violating our assumption of non-integer $t$). Hence, we conclude that the denominator of (\ref{eq: r}) is strictly positive.

Next, the numerator of (\ref{eq: r}) is strictly less than the denominator of (\ref{eq: r}), which is shown as follows. First, we can see that
\begin{align}
\mu[n]\Big(\lceil t \rceil - t\Big) &\buildrel{(a)} \over \leq \mu[n] - \Big [ \mu[n] - (\lceil t \rceil - t)\Big]^+ = \mu[n] - m_2[n],
\end{align}
where (a) is because $\mu[n] \leq 1$ and $\lceil t \rceil - t < 1$. Hence, we obtain
\begin{align}
\lfloor t \rfloor (\lceil t \rceil - t) &< t(\lceil t \rceil - t) =\sum_{n=1}^{N}\mu[n](t-\lfloor t\rfloor) \leq \sum_{n=1}^{N}\left(\mu[n] - m_2[n]\right) = t - \sum_{n=1}^{N} m_2[n],
\end{align}
which implies the numerator of (\ref{eq: r}) is strictly less than the denominator of  (\ref{eq: r}).

Finally, we need to show that the numerator of (\ref{eq: r}) is non-negative. Let $v\in\mathbb{Z}^+$ be the number of storage requirements which are greater than or equal to $t-\lfloor t\rfloor$,
\be
\label{eq: v}
v = \sum_{n=1}^{N}\mathbbm{1}\left ( \mu[n] \geq t-\lfloor t\rfloor  \right ).
\ee
Given (\ref{eq: v}), we establish two upper bounds on $\sum_{n=1}^{N} m_1[n]$. The first is given by
\be \label{eq: bound1}
\sum_{n=1}^{N} m_1[n] \leq v(\lceil t \rceil - t).
\ee
This holds because for any $n$ such that $\mu[n] \geq t-\lfloor t\rfloor$, we have
\be
m_1[n] =  \mu[n] - ( t - \lfloor t \rfloor)  \leq 1- ( t - \lfloor t \rfloor) = \lceil t \rceil - t
\ee
and there are $v$ such $n$'s. For any other $n$ such that $\mu[n] < t - \lfloor  t \rfloor$, we have $m_1[n] = 0$, which does not contribute to (\ref{eq: bound1}).
and, furthermore, the $N-v$ storage requirements which are less than $t-\lfloor t\rfloor$ can be ignored.
The second upper bound of $\sum_{n=1}^{N} m_1[n]$ is given by
\be \label{eq: bound2}
\sum_{n=1}^{N} m_1[n] \leq t - v(t- \lfloor t \rfloor).
\ee
This holds because the cumulative storage requirements of these $v$ DBs cannot exceed $t$. It can be shown that when $v<t$, (\ref{eq: bound1}) is a tighter bound, and when $v>t$, (\ref{eq: bound2}) is a tighter bound.\footnote{Note that, $v$ is an integer and $t$ is assumed to be a non-integer, therefore the case of $t = v$ is not valid.} Then by finding the integer $v$ in each region which gives the largest bound, we find
\be
\sum_{n=1}^{N} m_1[n] \leq \lfloor t \rfloor(\lceil t \rceil - t), \text{ for } v<t
\ee
and
\be
\sum_{n=1}^{N} m_1[n] \leq t - \lceil t \rceil(t- \lfloor t \rfloor), \text{ for } v>t.
\ee
Then, since $\lfloor t \rfloor(\lceil t \rceil - t) = t - \lceil t \rceil(t- \lfloor t \rfloor)$, for general $v$, we conclude that
\be
\sum_{n=1}^{N} m_1[n]\leq \lfloor t \rfloor(\lceil t \rceil - t)
\ee and the numerator of (\ref{eq: r}) is non-negative. Therefore, we have shown that $0\leq r < 1$ and this completes the proof of Lemma \ref{cl: 4}.

\bibliographystyle{IEEEbib}
\bibliography{references_d2d}

\end{document}